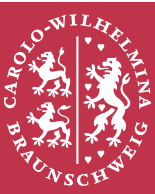

Technische
Universität
Braunschweig

Institute of
System Security

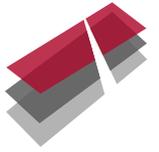

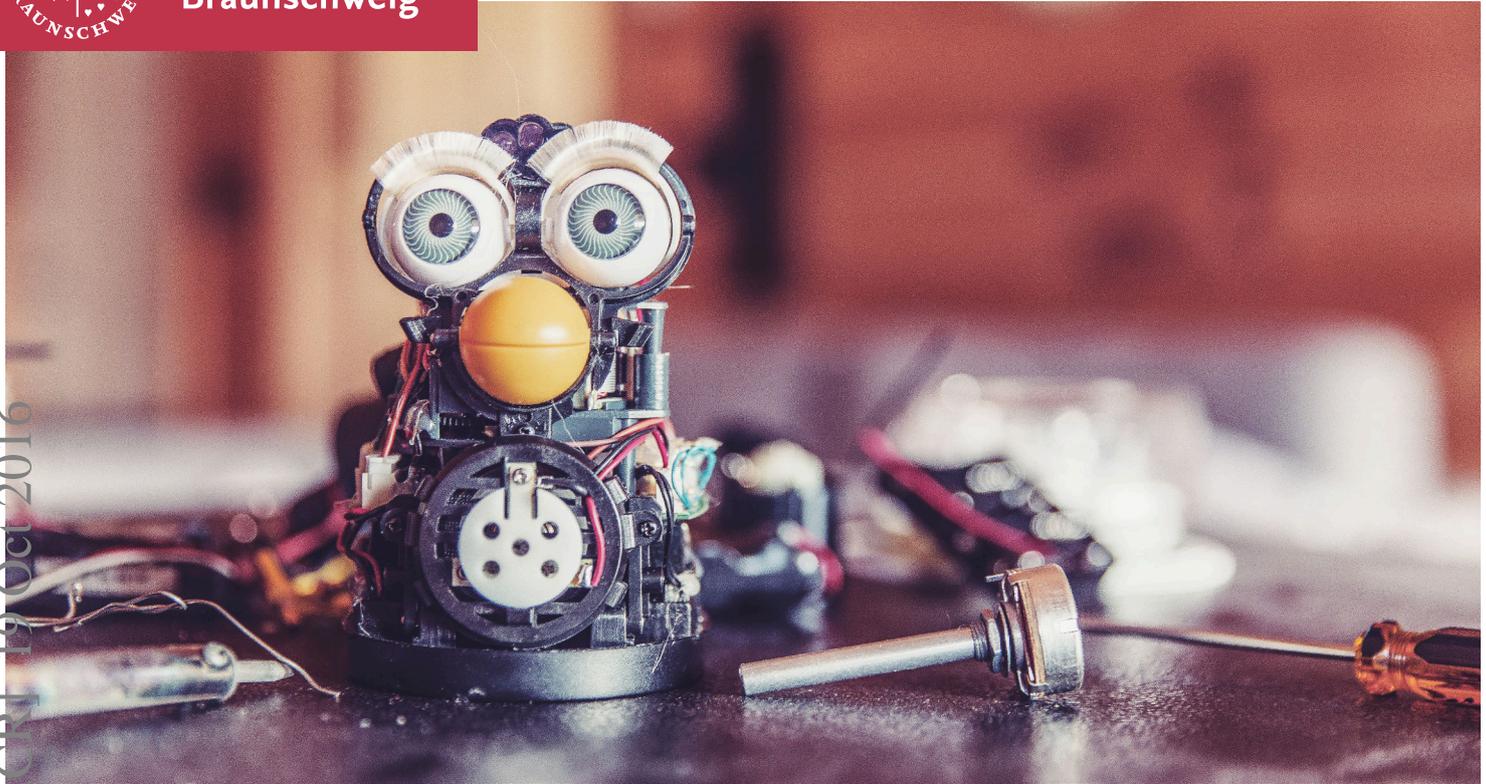

# From Malware Signatures to Anti-Virus Assisted Attacks

Christian Wressnegger, Kevin Freeman,
Fabian Yamaguchi, and Konrad Rieck

Technische Universität Braunschweig
Institute of System Security
Rebenring 56
38106 Braunschweig, Germany

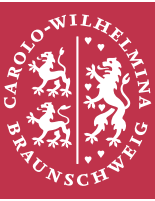

Technische
Universität
Braunschweig

## Abstract

Although anti-virus software has significantly evolved over the last decade, classic signature matching based on byte patterns is still a prevalent concept for identifying security threats. Anti-virus signatures are a simple and fast detection mechanism that can complement more sophisticated analysis strategies. However, if signatures are not designed with care, they can turn from a defensive mechanism into an instrument of attack. In this paper, we present a novel method for automatically deriving signatures from anti-virus software and demonstrate how the extracted signatures can be used to attack sensible data with the aid of the virus scanner itself. We study the practicability of our approach using four commercial products and exemplarily discuss a novel attack vector made possible by insufficiently designed signatures. Our research indicates that there is an urgent need to improve pattern-based signatures if used in anti-virus software and to pursue alternative detection approaches in such products.

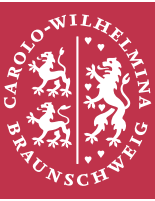

# 1 Introduction

Virus scanners are one of the most common defenses against security threats—despite well-known weaknesses and shortcomings. Millions of end hosts run these scanners on a regular basis to check files for infections and detect malicious code. Individuals, companies and even government organizations employ anti-virus software for fending off attacks at their desktop systems. The success and prevalence of these products largely build on their simple yet reliable functionality: Files are matched against a database of known detection patterns (signatures) which is regularly updated by the vendor to account for novel threats. Such pattern matching can be implemented very efficiently and is able spot all sorts of threats if appropriate and up-to-date signatures are available [see 3, 42].

Signature-based detection suffers from a well-known drawback: Unknown threats for which no signatures exist can easily bypass the detection. This problem is further aggravated by the frequent use of obfuscation in malicious code that obstructs static signature matching [10, 24, 27]. As a consequence, a large body of research has focused on developing alternative approaches for analyzing and detecting malicious software, for example, using semantic models [e.g., 7, 8], behavioral analysis [e.g., 9, 19, 22] or network traffic monitoring [e.g., 13, 14, 38]. Over the last decade, anti-virus vendors have increasingly adopted these detection mechanisms to keep up with malware evolution. Still, signatures based on byte patterns remain an integral part of security products and complement more sophisticated detection mechanisms.

Attackers are well aware of the prevalence of virus scanners and the detection mechanisms they employ. Thus, an adversary might not only seek means for evasion but take advantage of their presence. Anti-virus products have been shown to contain exploitable vulnerabilities as any other piece of software [e.g., 17, 30–34] and, according to leaked internal documents, even the NSA and GHCQ have set their hands on anti-virus software to infiltrate networks [11]. Alternatively, an attacker could also gain access to the deployed signatures and make use of them in an adversarial manner, for example, by injecting signatures into benign content.

In this paper, we focus on the latter scenario and explore the feasibility of such *anti-virus assisted attacks*. We introduce a novel method for automatically deriving signatures from commercial virus scanners which is agnostic to the implementation and provides an adversary with byte patterns that approximate the original signatures. With these patterns at hand, the attacker can draw a virus scanner's attention to benign data and flag chosen content as malicious, and thereby selectively block access or delete files. We show that due to inadequately designed signatures, this can be achieved by a remote attacker *without* direct access to the target system and data, or the availability of software exploits.

To assess the feasibility of this attack in practice, we apply our method for signature derivation to four commercial anti-virus products in an empirical study. We find that on average 38% of the derived signatures can be approximated by simple combinations of byte patterns. Several of these patterns match text strings, artifacts of packers or environment checks, and are mostly unrelated to the semantics of the considered malicious code. Furthermore our study shows that only 8% of such signatures match patterns that are detected by more than one anti-virus product considered in our experiments, enabling an attacker to play off gateway and end-user security solutions against each other by crafting target-specific inputs. We investigate the threat of using such pattern-based signatures as *malicious markers* to attack the availability of benign data and demonstrate the feasibility of such attacks in three different scenarios: 1) covering up password guessing, 2) deleting a user's emails and 3) facilitating web-based attacks by removing browser cookies. All three attack scenarios share the characteristic of having a virus scanner as privileged ally and remotely instrumenting it to delete or block user data.

In summary we make the following contributions:

- **Automatically deriving signatures.** We present a novel method for automatically deriving pattern-based signatures from virus scanners without the need for reverse engineering the software.

- **Identification of inadequate signatures.** In an empirically study with four commercial anti-virus products, we identify overly simplistic signatures that build on short byte patterns of resource or code artifacts.

- **Anti-virus assisted attacks.** Based on the derived signatures, we introduce a new class of anti-virus assisted attacks that limit access to benign data and demonstrate their feasibility in different scenarios.

The remainder of the paper is structured as follows: In Section 2 we present a brief overview of anti-virus signatures. We introduce our method for deriving signatures in Section 3 and empirically study its application and results in Section 4. In Section 5 we then present novel attacks based on malicious markers. Limitations and related work are discussed in Section 6 and Section 7, respectively. Section 8 concludes the paper.

## 2 Anti-Virus Signatures

Anti-virus software comprises a wide range of different analysis and detection techniques, including batch processing, on-access analysis, behavioral blocking and scheduled updating of signatures. While the underlying analysis engines often build on sophisticated concepts, such as bytecode interpreters [30] and other Turing-complete representations [5], the actual signature matching typically boils down to three basic strategies: *byte patterns*, *hash sums*, and in the widest sense of the word, *heuristics*. In the following, we describe each of these strategies in more detail to set the scope for our approach of deriving signatures.

## 2.1 Signatures based on Byte Patterns

The most common strategy for signature-based detection is the use of *byte patterns*. These signatures contain one or more constant sequence of bytes, possibly separated by gaps of varying size, that need to match with the content of a file to trigger a detection. These signatures can be efficiently matched using classic string-processing techniques, such as the Aho-Corasick algorithm [1, 15]. To this end, sets of strings are represented as a Trie that serves as a finite state machine for determining matches. This representation allows to efficiently model wildcards, ranges and character classes, thereby providing capabilities similar to those of regular expressions.

*Example of byte patterns.* The open-source scanner *ClamAV* defines a simple format for representing byte patterns, including disjunctions (aa|bb), gaps {n-m}, wildcards '?' and character classes [41]. Figure 1(a) shows a simplified version of such a signature for the *Virut* malware family. While *ClamAV* can generally match arbitrary data, in case of the provided example the signature describes x86 code that corresponds to a simple decryption routine. Figure 1(b) shows one instance matched by the above signature.

```
(8a|86)0666(29|31)(c8|d0|d8|e8|f8)(86|88)0646
```

(a) *ClamAV* signature W32.*Virut.si*

```
1    8a 06      ; mov  al, byte ptr [esi]
2    66 31 e8   ; xor  ax, bp
3    88 06      ; mov  byte ptr [esi], al
4    46         ; inc  esi
```

(b) Corresponding x86 code snippet.

Figure 1: *ClamAV* signature for the *Virut* malware and the correspoding x86 code snippet.

Formally, pattern-based signatures simply are a subset of regular expressions, yet for the purpose of signature derivation, we choose a different formal description that focuses entirely on their appearance. We observe that pattern-based signatures are given by sequences of disjunctions over bytes, possibly separated by gaps of varying size. Expressing each disjunction as a set containing its alternatives, a signature becomes a sequence of sets. We can describe this formally by defining a symbolic alphabet $S = \mathcal{P}(\{0, \ldots, 255\})$, where $\mathcal{P}$ is the power set and corresponds to all possible subsets of byte values. Additionally we use $\star$ as a shortcut for $\{0, \ldots, 255\}$ to represent irrelevant bytes. Each word $w \in S^*$ of the corresponding language then already fully expresses the format of a signature. However, these words alone do not account for the varying sizes of gaps. We hence introduce two functions, $l : \{1, \ldots, |w|\} \to \mathbb{N}$ and $h : \{1, \ldots, |w|\} \to \mathbb{N}$ that assign a minimum and maximum number of repetitions to each of $w$'s symbols. A pattern-based signature is then simply given by the tuple $s = (w, l, h)$.

## 2.2 Signatures based on Hash Sums

A second strategy frequently implemented by anti-virus software is matching based on *hash sums*. To this end, hash sums over complete files or parts of files are calculated and stored in the signature database. Simple checksums such as *CRC32* or cryptographic hash functions such as *MD5* or *SHA1* are often used due to the availability of fast implementations in both software and hardware [16, 30, 41]. While hash collisions may theoretically result in false positives, in practice this is not an issue in this particular field of application, making it an attractive choice for many vendors.

In comparison to byte patterns, hash sums enable matching large regions in a file with a compact signature. This approach, however, does not allow for wildcard characters or gaps, and therefore provides no means to match largely similar files with a single signature. In consequence, individual hash-based signatures are required for even the slightest variations of known malware samples, and thus pattern-based signatures may better meet the space requirements in the long run after all.

*Example of hash sums.* The open-source scanner *ClamAV* enables to define hash-based signatures either as hash sum of the complete file or specifically for PE files over individual sections [41]. Figure 2 illustrates both types of hash sums for the malware *Kido*, where in the first case the length of the matching region is specified in the second parameter (162970) and in the latter case in the first parameter (81920). Note that the type of the hash function is derived by *ClamAV* based on the size of the hash sum or the database file it is stored in.

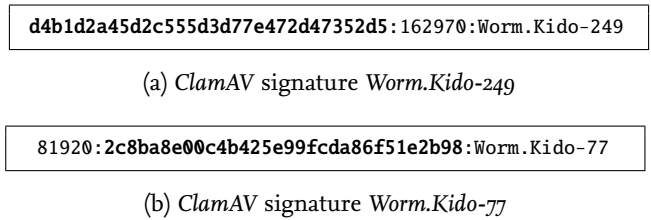

(a) *ClamAV* signature *Worm.Kido-249*

(b) *ClamAV* signature *Worm.Kido-77*

Figure 2: Example of two hash-based signatures, matching (a) the complete file and (b) a specific PE Section.

Conceptually, hash-based signatures can be interpreted as continuous byte patterns and thus the signatures can be described using the same formal description presented in Section 2.1. In particular, a hash sum with given offset and length can be represented as a continuous byte pattern that is preceded by a gap of fixed size.

## 2.3 Heuristics

Aside from byte patterns and hash sums, anti-virus software often employs several additional heuristics for detection of security threats. These heuristics include the inspection of instruction counts, the analysis of API usage, n-gram detection models, and in some cases even machine learning techniques. In comparison to signatures based on byte patterns and hash sums, these detection approaches are often bound to a concrete execution context and can capture complex semantics, such as different events necessary to trigger an infection. Due to this complexity heuristics are less suitable for constructing malicious markers and we thus put our focus on pattern-based signatures only.

# 3  Deriving Malware Signatures

A classic approach for deriving signatures from virus scanners is reverse engineering the respective analysis engines and dumping their signature databases [30]. Such reverse engineering, however, relies on tedious manual work and needs to be repeated for each individual scanner. We in contrast introduce a simple and intuitive, yet generic method that is agnostic to the inspected virus scanner: Given a set of known malware samples, we strategically create modifications of these files and scan them to see whether the scanner still flags them as malicious. This procedure allows us to derive a signature by piecing together bytes that are observed to be relevant for detection over different samples and runs. Although simple in design this method is sufficient to obtain suitable markers for anti-virus assisted attacks as introduced in Section 5.

Our method for the derivation of signatures executes the following three steps which can be applied to any possible virus scanner:

1. *Detecting relevant bytes.* First, we determine relevant bytes in each malware sample by utilizing feedback from the virus scanner over multiple runs (Section 3.1).

2. *Sequence alignment and merging.* We proceed to align the relevant bytes from samples with the same signature and merge them into a single sequence (Section 3.2).

3. *Creation of signatures.* Finally, we transform the merged sequences into a valid signature format, yielding the final signature (Section 3.3).

In the following, we discuss each of these steps in detail and describe the optimizations we perform compared to a naive implementation.

## 3.1  Detecting Relevant Bytes

We begin our analysis by processing each malware sample independently to derive bytes *relevant* for detection. To achieve this, the main idea is to simply flip—bitwise negate—one byte after another and run the target virus scanner on the resulting file. If the modified file does not trigger an alarm, the changed byte is relevant for detection and thus we include the original value of that byte in our signature. If in contrast, the scanner shows no reaction to our modification, the byte seems to be irrelevant.

There are three problems with this approach in practice. First, determining the exact signature requires to exhaustively test all possible combinations rather than simply flipping bytes. Second, the largest portion of a virus scanner's runtime is taken up by its initialization. Passing each file variation to the scanner separately therefore induces a high runtime overhead. Third, a naive implementation of this approach requires quadratic disk space and its runtime is dominated by the disk's I/O operations.

As a remedy, we restrict the derivation process to signatures that match all samples of the same malware family in our dataset but are not necessarily complete. For the application as malicious markers in anti-virus assisted attacks this is a sufficiently good approximation. Moreover, we reduce the runtime overhead induced by the scanner's initialization by passing samples to the scanner in large batches rather than separately.

We finally address the third problem by formulating the following algorithm that follows a divide-and-conquer approach to perform byte flipping: The complete file is first divided into $k$ partitions. For each partition, all bytes are flipped and the virus scanner is applied to the modified file. If a partition does not contain relevant bytes (the scanner still detects the malware) no further inspection is needed and the complete partition is considered to be irrelevant for the signature. Otherwise, the same procedure is recursively applied until partitions can no longer be divided. In effect, large portions of a file can be quickly marked as irrelevant and dismissed.

The procedure presented thus far is well suited for signatures based on byte patterns but inefficient when dealing with hash-based signatures, particularly if the hash sum is calculated over the entire file. Our approach would flip all $n$ bytes individually and requires to scan the resulting $n$ files in order to derive this relation. To speed up this process we introduce two thresholds $t_h$ and $t_p$. The first relates to the ratio of bytes considered relevant during one iteration of the divide-and-conquer process. The second specifies the partition sizes for which our heuristic is applied: if $t_h = 99\%$ of a partition with at least $t_p = 25,000$ bytes are marked as relevant, we conclude that we are dealing with a hash-based signature and focus the search on the edges of the partition to determine the exact region the hash is calculated on. This allows us to decide whether or not we are dealing with a hash-based signature at an early stage and accelerate the process significantly.

As a result of this analysis, we obtain a preliminary signature $s = (w, l, h)$ for each sample $x$ composed of $m$ bytes $x_1, \ldots, x_m$. We construct this signature by first creating a format description $\hat{w}$ where $\hat{w}_i = \star$ if $x_i$ corresponds to an irrelevant byte, and $\hat{w}_i = \{x_i\}$ if $x_i$ is a relevant byte. We proceed to create a compressed form of $\hat{w}$ referred to as $w$ by scanning $\hat{w}$ from left to right and merging consecutive irrelevant bytes (gaps). Relevant bytes, however, are preserved for the sake of signature alignment as described in Section 3.2. We additionally store information about the length of the gaps as the minimum and maximum number of repetitions. That is, for each byte $w_i$, we set $l(i)$ and $h(i)$ to the number of symbols of $\hat{w}$ merged to obtain $w_i$. The resulting signature may still be changed in the next step to account for information contained in other samples tagged with the same signature label.

## 3.2 Sequence Alignment

We proceed by grouping malware samples according to the label assigned by the scanner. For a given group of corresponding signatures $X$, we then create a joint signature by employing the Needleman-Wunsch algorithm [29]. That is, we align their corresponding format descriptions given by the set $\{w \mid (w, l, h) \in X\}$. During this procedure we ignore the minimum and maximum number of repetitions encoded by the functions $l$ and $h$ at first and merely compare the signature's format descriptions.

Given two strings $v$ and $w$, the Needleman-Wunsch algorithm attempts to align these strings, that is, it creates new strings $\hat{v}$ and $\hat{w}$ from $v$ and $w$ respectively by introducing an arbitrary number of irrelevant bytes denoted as $\star$ between bytes of $v$ and $w$. These additional irrelevant bytes serve as padding, ensuring that $\hat{v}$ and $\hat{w}$ are of same size. The algorithm attempts to ensure that $\hat{w}_i$ is equal to $\hat{v}_i$ for as many of the strings' positions $i$ as possible. For merging the resulting alignment we combine the length specifications denoted by $l$ and $h$ and extend the bounds such that the range is maximized. Moreover, for positions $i$ where $\hat{v}_i$ and $\hat{w}_i$ are unequal, we simply merge the sets $\hat{v}_i$ and $\hat{w}_i$ by applying the union operator. To create a signature for an entire set of strings, we iteratively apply the Needleman-Wunsch algorithm to merge one signature at a time with the preliminary version of our final signature.

To further detail the range specifications one can employ a similar approach as used for identifying relevant bytes (see Section 3.1). We insert or delete dummy bytes at the locations of irrelevant bytes to determine the number of bytes that may occur in between sequences of relevant bytes and apply the virus scanner to it. To this end, we again make use of a (binary) divide-and-conquer strategy in order to speed up the process. While this approach provides a precise solution, this strategy is time consuming and thus we omit this step for our experiments and rely on the alignment described above.

As a result of this step, we obtain a signature for each group of samples, that is, a tuple $(w, l, h)$ where $w$ describes the signatures format as a sequence of disjunctions over bytes, and $l(i)$ and $h(i)$ are the minimum and maximum number of repetitions of the $i$-th symbol in $w$.

## 3.3 Creation of Signatures

We finally construct signatures compliant with those used by *ClamAV* and in accordance with our formal definition. To this end, we join the values of each symbol $w_i$ using the alternate operator '|' surrounded by brackets, which may be omitted if a symbol contains one value only. Additionally, we annotate this specification of a symbol with its minimum and maximum number of repetitions expressed as ranges {n-m}, where n and m are specified by a signature's functions $l$ and $h$. Strictly specified ranges such as {n-n} are simplified to {n}. Repetitions of exactly one, {1-1} and {1} respectively, are omitted for simplicity, yielding a final signature as shown in Figure 3.

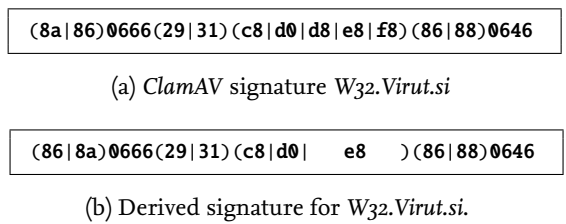

(a) *ClamAV* signature W32.*Virut.si*

(b) Derived signature for W32.*Virut.si*.

Figure 3: Signatures for the *Virut* malware (a) as specified in *ClamAV*'s database and (b) as derived by our method (missing bytes are indicated by spaces).

# 4 Empirical Study

Equipped with a method for automatically deriving signatures from virus scanners, we conduct an empirical study with four popular anti-virus products and the open-source scanner *ClamAV*. As we are not interested in comparing individual security vendors but in gaining insights into used signatures, we reference these scanners in the following as *AV1* to *AV5*, with *ClamAV* being *AV1* as our baseline. In particular, we carry out four experiments on a recent malware dataset that is detailed in Section 4.1:

- First, we demonstrate the viability of the proposed method for signature derivation in a controlled experiment using *ClamAV* (Section 4.2).

- Second, we apply our method to the virus scanners we do not have ground truth for and perform a quantitative study of the derived signatures (Section 4.3).

- Third, we investigate whether or not signatures from different vendors overlap, that is, cover identical malware bytes, and if so, to which extend (Section 4.4).

- Fourth, we study the quality of deployed signatures with special focus on semantics and whether or not they are bound to a specific context (Section 4.5).

## 4.1 Malware Dataset

The quality of derived signatures hinges on a representative dataset of malware for deriving corresponding signatures. We thus collect 9,969 malware samples that are detected by all five scanners considered in our evaluation. In particular, we have been given access to submissions to the Virus-Total service, allowing us to gather a broad selection of recently submitted files. A brief overview of the dataset is presented in Table 1. The vast majority of the files are applications or dynamic libraries in the *Portable Executable* format. The remaining files correspond to archives, Windows shortcuts and other carriers of malicious code. Depending on the applied virus scanner, the files in the dataset are assigned to roughly 250 malware families, where the concrete number of different signature labels assigned by the scanners ranges from 277 up to 1,327 (see the first column in Table 2).

| Type | # | Type | # |
|------|---|------|---|
| PE32 | 9,721 | Windows shortcuts | 21 |
| PE32+ | 38 | HTML/XML | 121 |
| MS-DOS executable | 38 | Text | 15 |
| Archive formats | 14 | Others | 1 |

Table 1: Overview of the evaluation dataset.

## 4.2 Controlled Experiment: ClamAV

In the first experiment, we evaluate our approach in a controlled setup using the open-source scanner *ClamAV* for which all signatures are publicly available and we thus have ground truth to assess the reconstruction capabilities of our approach. To this end, we first derive signatures for all samples in our datasets and then compare the output to the corresponding signatures of *ClamAV* using the string edit distance [23]. In this experiment we focus on static pattern-based signatures as returned by our method, but skip hash-based signatures as these are not suitable for being used as malicious markers (cf. Section 5). Moreover, we replace gaps in all signatures with a generic wildcard to account for minor differences in the gap ranges.

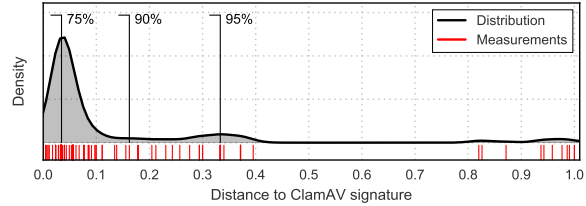

Figure 4: Quality of derived signatures, measured as the normalized edit distance between the derived signatures and the corresponding *ClamAV* signatures.

Figure 4 presents the results of this comparison as the distribution of the observed distances to the original signatures. To achieve a comparison between signatures of different lengths, we normalize the edit distance $d(x, y)$ between two signatures $x$ and $y$ as follows:

$$\tilde{d}(x, y) = \frac{d(x, y)}{\max(|x|, |y|)} \in [0, 1]$$

where $|x|$ and $|y|$ correspond to the lengths of the signatures. The resulting distribution of normalized distances has a sharp peak close to 0.0 and roughly 50% of all derived signatures match their counterpart in the *ClamAV* database almost perfectly. Moreover, 75% of the derived byte patterns differ from the original signature by at most 4% of the content and only a minor fraction cannot be correctly derived by our method.

As an example of this experiment, let us re-consider the derived signature shown in Figure 3. The two expressions are almost identical except for the third disjunction, where the derived signature misses two out of five possible bytes (indicated as spaces). As shown in Figure 1, this signature corresponds to a simple decryption loop that is varied by the malware authors using different instructions and registers. Our dataset comprises samples of *Virut* with three of the five variants encoded in the *ClamAV* signature and thus the derived signature matches their counterpart not exactly. Still, the rest of the byte patterns and their disjunctions are precisely uncovered by our approach.

In addition to the successful derivations, we also inspect cases where our method fails to arrive at a good approximation of the original signature. One example is a variant of the malware family *Adware.Somoto*[1] which *ClamAV* identifies using a compact byte pattern that our approach misses. A closer investigation reveals that the corresponding sample is a self-extracting archive of the *Nullsoft Scriptable Install System* (NSIS) for which *ClamAV* introduced an unpacker in version 0.91. Dynamic analyses and unpacking are out of scope of our method.

---

[1]md5: `e7f7d72e1f7f3371e2cf495d76b2b88a`

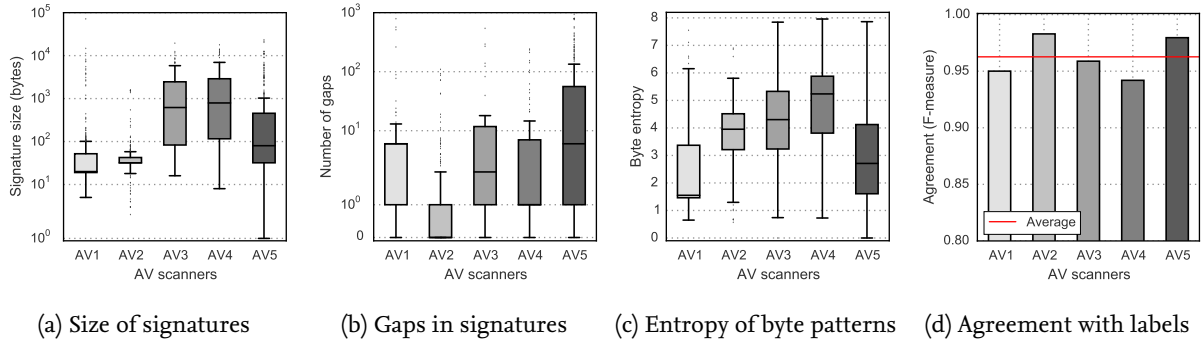

(a) Size of signatures    (b) Gaps in signatures    (c) Entropy of byte patterns    (d) Agreement with labels

Figure 5: Statistics for the derived byte-pattern signatures. The agreement with anti-virus labels is given as F-measure of the labels and the clustered signatures.

## 4.3 Quantitative Study: AV2–AV5

We proceed to derive signatures for the remaining four commercial anti-virus products. The setup is kept identical to the previous experiment with the only exception that we do not have access to the ground truth for the original signatures. We therefore focus on statistically assessing the quality and content of the derived signatures.

|  | **Derived Signatures** | | | | | | |
|---|---|---|---|---|---|---|---|
|  | #Labels | Hash sums | | Byte patterns | | Unknown | |
| **AV1** | 1327 | 940 | (71%) | 377 | (28%) | 10 | (1%) |
| **AV2** | 277 | 1 | (0%) | 255 | (92%) | 21 | (8%) |
| **AV3** | 522 | 323 | (62%) | 69 | (13%) | 130 | (25%) |
| **AV4** | 282 | 178 | (63%) | 93 | (33%) | 11 | (4%) |
| **AV5** | 586 | 177 | (30%) | 353 | (60%) | 56 | (10%) |
| Average | 598 | 323 | (54%) | 229 | (38%) | 45 | (8%) |

Table 2: Results of signature derivation.

Table 2 provides a first overview of the derived signatures, where *AV1* corresponds to *ClamAV* and the remaining scanners to commercial products. The scanners considerably differ in type and amount of signatures that can be derived. On average, our approach is able to infer 38% pattern-based signatures and 54% hash-based signatures, whereas only the remaining 8% cannot be uncovered.

*AV2* especially stands out with 92% byte patterns and only a single hash-based signature, while for the other scanners at least one third of all signatures corresponds to hash sums. On the other end of the scale, for *AV3* we fail to extract one forth of all signatures. This may indicate the use of advanced heuristics or that multiple pattern-based signatures are assigned to the same label, implicitly constructing a signature with long disjoint patterns. We discuss this in more detail in Section 6.

To illustrate the quality of the derived signatures, let us consider the byte patterns shown in Figure 6 and 7, respectively. Figure 6 shows a signature for the malware family *Swizzor*. Clearly, an effort has been made to pinpoint the decryption loop of the malware (starting at line 7) that unveils the malicious payload at runtime. In particular a 5-byte long key is used to decrypt data using the XOR operator (line 8).

```
1   83 c7 43          ; add  edi, 0x43
2   89 fa             ; mov  edx, edi
3   83 ea 2e          ; sub  edx, 0x2e
4   {2}               ; 2-byte gap
5   00 00
6   33 c0             ; xor  eax, eax
7   8a 1f             ; mov  bl, byte ptr [edi]
8   32 1C 10          ; xor  bl, byte ptr [eax + edx]
9   88 1F             ; mov  byte ptr [edi], bl
10  40                ; inc  eax
11  83 f8 05          ; cmp  eax, 5
12  7c 02             ; jl   +2
13  33 c0             ; xor  eax, eax
14  47                ; inc  edi
```

Figure 6: Derived signature for the *Swizzor* malware: The byte patterns match a XOR-based decryption loop (line 7) with a 5-byte key (line 11).

By contrast, the signature shown in Figure 7 contains an ASCII string that corresponds to the Windows product key of the Anubis sandbox. The instructions at lines 6–11 are part of a call to the `RegOpenKeyExA` API function. This signature captures a simple environment check used by several backdoors that compares the product key of the execution environment against known sandbox systems.

```
1   49 64 00 00             ; "Id"
2   37 36 34 38 37 2d       ; "76487-337-8429955-22614"
3   33 33 37 2d 38 34
4   32 39 39 35 35 2d
5   32 32 36 31 34 00
6   50                      ; push eax
7   81 c4 f4 fe ff ff       ; add esp, 0xfffffef4
8   33 c0                   ; xor eax, eax
9   54                      ; push esp
10  6a 01                   ; push 1
11  6a 00                   ; push 0
```

Figure 7: Derived signature for a generic backdoor: The byte patterns correspond to the Windows product key of the Anubis sandbox (line 2–5).

Not all of the signatures we derive can be interpreted as clearly as these examples. Many signatures match resources, wrongly aligned code, import tables or even compressed data. To get a better understanding for the quality of these signatures, we statistically analyze the characteristics of the corresponding byte patterns derived by our method. Figures 5(a) to 5(c) show for each of the five scanners the distribution of the signature size, the number of gaps and the entropy of the patterns. With respect to the size, *AV2* again draws attention: It appears that this scanner makes use of very compact and equally sized patterns that contain no or only a few gaps. While short signatures generalize well, they may also induce more false positives which presumably is the reason why other scanners such as *AV3–AV5* use longer signatures—partly also with more gaps between the byte patterns. As demonstrated in Section 5 short signatures are good candidates for malicious markers.

As malicious markers might be restricted to a specific character set, it is also interesting to see how the entropy of the byte patterns is distributed. An entropy of 0.0 indicates uniform byte patterns, such as paddings, while values close to 8.0 are reached by random, encrypted or compressed data. Matched import tables reside on the lower end of the scale due to the identical significant bytes of addresses and x86 code is mostly located in the middle and upper half of the scale. Three of the scanners cover almost the full entropy range, indicating the diversity of the byte patterns in the signatures.

Finally, we inspect how well the pattern-based signatures have been derived by our method. Since we do not have ground truth, we rate the quality of the signatures by clustering them and comparing the clusters to the groups of labels the signatures correspond to. This conformity is plotted in Figure 5(d) as the F-measure for the individual anti-virus products. On average the derived signatures yield an agreement with the labels of 96%, that is, for most of the dataset, similar signatures are assigned to the same label, whereas dissimilar signatures are associated with unequal labels. While this experiment cannot prove that the derived signatures exactly match their counterparts, it demonstrates that we infer a representation that is at least closely related to the original signatures.

## 4.4 Signature Overlap

Although anti-virus vendors openly share samples, concrete signatures usually are not exchanged as these represent a company's trade secret. We investigate to which extend signatures nevertheless overlap between competing anti-virus products. To this end we measure the number of (half-)bytes in relation to a signature's length that overlap between products. Figure 8 shows the average overlap of the five considered engines to each other. Only 5.2% of the derived signatures *fully* overlap with the signature of another vendor in a way that both products raise an alarm on the same substring of bytes. Consequently, for 94.8% of the signatures a virus scanner raises an alarm while others do not. 21.9% even have no resemblance among different anti-virus products at all.

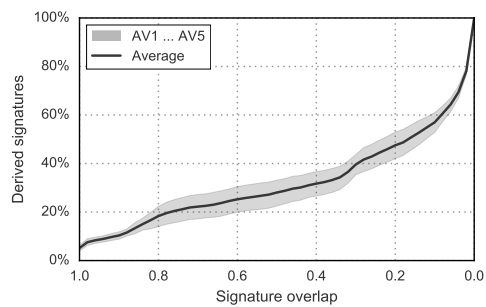

Figure 8: Average overlap (ratio of identical bytes) of signatures derived from *AV1–AV5*.

## 4.5 Context and Semantics of Signatures

As final experiment of our study, we analyze the quality of the derived signatures. We are interested in determining whether the signatures are bound to a specific context within the file and if they model semantics of the targeted malicious code. To this end, we examine how well the derived signatures can be implanted in benign files that have nothing in common with the original malware. Since our dataset is mainly comprised of PE executables, we consider a set of benign applications of the same format for this task, which we have verified not to be flagged as malicious by any of the scanners. In particular, we choose the following applications from the Windows system directory: *Calculator, Windows Command Prompt, Internet Explorer, Microsoft Management Console,* the *System Configuration Application* and *Microsoft Paint.*

We proceed to implant the signatures derived by our method into these applications, irrespective of the relative position to PE sections or existing code, and apply the virus scanners to the resulting files. A breakdown of the detection results is provided in Table 3. On average 68% of the implants are successful. Apparently, several of the pattern-based signatures are applied without checking the context or semantics of the matching region for plausibility, allowing the direct transfer of signatures from one file to another. The situation is especially troublesome for *AV1, AV2* and *AV3* that simply flag the majority of implants as malicious. *AV5* appears to be the only one to use semantics-aware matching in most of the cases. Still, the overall quantity of working implants is higher than for *AV3* or *AV4*.

|        | Byte patterns | Implantable     |
|--------|---------------|-----------------|
| **AV1**  | 377           | 358  (94.96%)   |
| **AV2**  | 255           | 227  (89.02%)   |
| **AV3**  | 69            | 65  (94.20%)    |
| **AV4**  | 93            | 54  (58.06%)    |
| **AV5**  | 353           | 80  (22.66%)    |
| Average | 229           | 157  (68.35%)   |

Table 3: Detection results on pattern-based signatures implanted in benign PE files.

# 5 Anti-Virus Assisted Attacks

Over the last years, numerous software flaws have been identified in popular anti-virus products [e.g., 17, 30]. For example, Ormandy recently disclosed vulnerabilities in products of ESET [31], Kaspersky [32], AVG [34] and FireEye [33]. These findings are especially critical as the corresponding products are employed in many network and host defenses. According to leaked internal documents, even the NSA and GHCQ have set their hands on anti-virus software to infiltrate networks [11].

We advance this line of work and present a novel class of anti-virus assisted attacks, called *malicious markers*, that does not rely on exploiting vulnerabilities but is based on the weak design of pattern-based signatures. Although modern anti-virus products comprise several different detection mechanisms, signatures based on byte patterns are still widely employed and our attack affects all products that operate on poorly designed patterns. In the following, we introduce malicious markers as attack vector in detail (Section 5.1) and demonstrate the effect of these attacks in different cases studies (Section 5.2).

| | Families | Derived signatures | *ASCII characters* * | *Printable-1* | *Printable-2* | *Printable-3* * | *Printable-4* * |
|---|---|---|---|---|---|---|---|
| **AV1** | 37 | 35 | 34 | 34 | 20 | 7 | 11 |
| **AV2** | 28 | 27 | 26 | 26 | 12 | 3 | 4 |
| **AV3** | 28 | 27 | 27 | 27 | 8 | 2 | 5 |
| **AV4** | 37 | 33 | 31 | 31 | 3 | 1 | 0 |
| **AV5** | 36 | 31 | 31 | 31 | 12 | 6 | 5 |

Table 4: Signatures derived from 200 PHP scripts partitioned by used characters sets: *ASCII* (\x01-\x7F), *Printable-1* (all printable ASCII characters), *Printable-2* (\x20-\x7F), *Printable-3* (*Printable-2* excl. = and ;) and *Printable-4* (*Printable-2* excl. slashes and :). Sets marked with * have been used for the attacks.

## 5.1 Attack Vector

*Malicious markers* rely on implanting malware signatures in benign data. If the virus scanner does not consider additional context for its decision, the tampered data is flagged as malicious and access is blocked by quarantining or even deleting the underlying file. At a first sight this attack scenario seems of little impact at best: It is obvious that an attacker can trigger alerts by attaching malware to benign files and there is nothing wrong with flagging such tampered data as malicious. Moreover, the attacker needs to be able to write to files, which would give rise to several other and more powerful attacks.

However, we consider a more subtle setting where a remote attacker can only implant a couple of carefully selected byte patterns into network communication. First, this is a much more realistic scenario and, second, smuggling through malicious markers suddenly becomes an appealing effort: By sending specially crafted network data that eventually is stored in a file, an attacker can instrument a virus scanner to block or delete this file without direct access. This, of course, requires the anti-virus software to be configured for automatically quarantining detected files, which fortunately is the default setup in several anti-virus products.

As reported in Section 4, many signatures from the dataset used in the empirical study are already implantable into benign files as is. Yet, similar to shellcodes in network attacks, useful malicious markers need to consist of certain bytes to pass input validation of common network protocols and end-user applications, such as ASCII or printable characters. Although signatures for executable files may happen to match this criteria by chance, other file types of malware promise to yield a larger amount of signatures usable as malicious markers.

To test our attack, we thus additionally consider 200 malicious PHP files detected by all five virus scanners. Depending on the applied virus scanner the samples correspond to 28–37 malware families. For each family, we derive signatures and inspect their suitability for the use as malicious markers. To this end, we categorize them into different character sets as shown in Table 4. Although we are only operating on 30 to 40 different signatures, we are able to find suitable malicious markers for all virus scanners and character sets. The actual effort for an adversary to find suitable markers for a given network application is thus rather low.

## 5.2 Case Studies

We proceed to demonstrate the threat and feasibility of implanting signatures in network data in three different scenarios: 1) covering up password guessing, 2) deleting a user's emails and 3) facilitating web-based attacks by removing browser cookies. For each scenario we choose a different anti-virus product to demonstrate the generality of the approach.

*No logs, no crime.* Similar to the empirical study presented earlier, we begin by considering the open-source scanner *ClamAV* as an attack target. We aim at removing traces of password guessing, for example, over SSH, POP3 or IMAP. For this purpose, we exploit that Linux and other UNIX systems keep track of failed login attempts in a log file, often called `auth.log` as show in Figure 10. Note that the username which is controlled by the attacker is stored in clear in the log file. It turns out that a large range of printable characters (*Printable-4*) can be used as username and is written *verbatim* to the file.

As a consequence, an attacker may finish each iteration over a list of guessed passwords with a set of malicious markers, i.e., specially crafted login names that correspond to anti-virus signatures. If the attacked host is running a virus scanner configured to delete or quarantine viruses, any file containing such a malicious marker is deleted or at least moved to a different location. This not only makes manual investigation of the attack hard but may also inhibit the functionality of tools analyzing log files to stop password guessing, such as *fail2ban* [18].

```
if(function_exists('exec'))@exec(if(function_exists(
'shell_exec'))@shell_exec(if(function_exists('system'))
@system(if(function_exists('passthru'))@passthru(
```

Figure 9: *ClamAV* signature labeled `PHP.ShellExec`

Figure 9 exemplarily shows a signature used by *ClamAV* that is accepted as login name and allows to tag the authentication log as malicious. Once the SSH daemon writes the malicious marker to `auth.log` *ClamAV* steps in to remove the file and destroys all evidence of the previously attempted attack. With the same technique any log file can be deleted that stores received network data in clear. Other imaginable targets are, for instance, log files of web, name and database servers that record requests and queries verbatim.

```
Feb  2 23:59:15 alice sshd[6126]: Invalid user  mallory  from 111.202.98.106
Feb  2 23:59:15 alice sshd[6126]: input_userauth_request: invalid user  mallory  [preauth]
Feb  2 23:59:17 alice sshd[6126]: pam_unix(sshd:auth): check pass; user unknown
Feb  2 23:59:17 alice sshd[6126]: pam_unix(sshd:auth): authentication failure;
                                  logname= uid=0 euid=0 tty=ssh ruser= rhost=111.202.98.106
Feb  2 23:59:18 alice sshd[6126]: Failed password for invalid user  mallory  from 111.202.98.106 port 46447 ssh2
Feb  2 23:59:19 alice sshd[6126]: Connection closed by 111.202.98.106 [preauth]
```

Figure 10: Excerpt of Linux's `auth.log` showing a failed attempt of `mallory` signing into a host called `alice`.

*Deletion of emails.* With a slight twist, malicious markers can also be used to obstruct the delivery of emails. To illustrate this setting, we consider the commercial anti-virus product *AV2* operated on Windows. As target we choose *Mozilla Thunderbird*, which stores the user's emails in a variation of the mbox format family [4]. While attachments and binary data are encoded, the email body is stored verbatim in this format. This enables an attacker to smuggle in malicious markers by sending crafted emails. The adversary is free to use any ASCII encoded characters, including non-printable, but excluding ASCII extensions and the NUL-character (*Printable-1*). Figure 11 shows a suitable candidate for *AV2* that can be used as implant in this setting.

It suffices that the attacker delivers a single email to the victim to trigger quarantining or deleting the inbox database. The crafted email does not even need to look suspicious, as the attacker may use ASCII control characters, such as \f (NP form feed, new page) or a sequence of newline characters or whitespaces, to hide the malicious marker from being displayed in clear sight in *Mozilla Thunderbird* and other mail clients. Note that it is not possible to simply use a complete malware binary for this attack as it likely contains non-printable characters and thus is incorrectly stored in the email database. Moreover, chances are high that the malware binary would be filtered out by the email gateway already, whereby the malicious marker most probably would not. We discuss this in more detail in Section 5.3.

```
=\"s\"+\"p\"+\"li\"+\"t\";
```

Figure 11: *AV2* signature labeled `JS:Decode-BHU [Trj]`

*Removing browser cookies.* The presented attack vector can also be adapted to assist in web-based attacks. In this scenario our goal is to force a user to re-login into a web application and thereby enable tampering with the authentication process. For this setting, we use *Google Chrome* as the target application and consider *AV5* as our partner in crime. Google Chrome stores its cookies in a simple SQLite3 database called `User Data\Default\Cookies`. Interestingly, while the actual value of the cookie is strictly encoded when stored in this database, the name of the cookie is not: A cookie name may hence consist out of any printable character in the range from \x20 to \x7E, except for semicolons and the equality sign (*Printable-3*).

An attacker may hence simply provide a specially crafted cookie, with the malicious marker as its name, on a website and lure its victim into visiting it (no evolved web-based attack, e.g., MITM, needed). On access of the webpage the anti-virus product blocks access to the cookie database, thereby forcing the user to re-authenticate with certain web applications. Figure 12 exemplarily shows a signature used by *AV5* that meets the mentioned criteria of allowed characters.

The same attack vector can be used to tamper with any data that contains user-controlled information and is stored in files by the browser. *Mozilla Firefox*, for instance, also stores *HTML5 local storage objects* in a SQLite database and thus is also a suitable target for malicious markers in web-based communication.

```
\x65\x76\x61\x6C\x28\x67\x7A\x69\x6E\x66\x6C\x61\x74\x65
\x28\x62\x61\x73\x65\x36\x34\x5F\x64\x65\x63\x6F\x64\x65
\x28'7X1re9s2z/Dn9VcwmjfZq+PYTtu7s2MnaQ5t2jTpcugp6ePJsmx
rkS1PkuNkWf77C4CkREqy43S
```

(a) Originally as matched by *AV5*

```
eval(gzinflate(base64_decode('7X1re9s2z/Dn9VcwmjfZq+PYTt
u72MnaQ5t2jTpcugp6ePJsmxrkS1PkuNkWf77C4CkREqy43S
```

(b) Same signature with resolved encoding.

Figure 12: *AV5* signature labeled `Backdoor.PHP.ASQ`

## 5.3 Mitigations

The presented malicious markers can be mitigated on different stages of the attack path, ranging from the network transmission to the affected applications and exploited anti-virus products.

*Network-based mitigation.* The transmission of implants can be effectively blocked, if the same signatures that are used on client systems are also deployed on a network gateway, for example as part of an intrusion detection system. In this setting, tampered cookies, emails and logins can be filtered out before reaching the client systems. This requires all security products on the host and network level to originate from the same vendor, as signatures usually are not shared among vendors. However, reliably enforcing such a homogeneous network setup is difficult and might even be impossible in some scenarios.

*Application-based mitigation.* The presented attack exploits the fact that several applications write data retrieved over the network verbatim to a single file. A simple defense strategy is thus to isolate content originating from different senders. For example, *Microsoft Edge* stores cookies in separate files rather than in a common database, while *Apple Mail* writes each received email to an individual file. In some settings, however, it is not feasible to separate content or determine its origin. For example, storing each entry of an authentication log in a separate file is far from being a practical solution.

An alternative strategy is to encrypt retrieved data with a local key before storing it to a file. This encryption renders any attack using malicious marker ineffective, at the prize of obstructing direct access to the stored content. A less effective but more practical variant might be to simply encode or compress the stored content, which definitely raises the bar for the attacker but does not rule out malicious markers in general, as AV products occasionally also match encoded or compressed artifacts as illustrated in Figure 12.

*Improving anti-virus products.* A prerequisite for the success of malicious markers is the interplay with an anti-virus product operated on a victim's system. Hence, there exist different options for countering the attacks by adapting anti-virus products. Quick and easy solutions are, for example, the blacklisting of files for quarantining and the binding of signatures to certain file types.

In the long run a more effective approach is to eliminate the feasibility of implanting signatures: This, however, requires pattern-based signatures to also account for the semantics and context of malicious code. It is not sufficient to simply spot the appearance of a pattern, but rather ensure that it is correctly embedded in the context of malicious code—a challenging yet necessary task to improve anti-virus products.

## 6  Limitations

Anti-virus assisted attacks using malicious markers target a specific albeit wide-spread type of signatures. Naturally, this approach is thus subject to limitations which we discuss in the following.

*Heuristics and dynamic execution.* As stated in Section 3, we do not consider any detection mechanisms based on heuristics. In particular, this includes all signature matching approaches that are based on the results of dynamic execution or unpacking. Although the underlying matching techniques might be similar but applied to execution logs or memory dumps, we do not consider the underlying patterns for our attacks. Note that injecting patterns from dynamic execution, such as sequences of system calls, into a benign process is a difficult problem and beyond the scope of this work.

*Alternative signatures.* In practice, virus scanners often use different signatures for the same malware family, such as *W32.Virut.a* and *W32.Virut.b*. However, it is also possible that multiple signatures map to the same label. Effectively, this can be thought as a disjunction of signatures. If these individual signatures match in different malware samples, we are able to derive the disjunction correctly. However, a problem arises whenever several of such signatures occur in the same file. If these are disjoint, our method is not able to derive any of them, since whenever the byte sequence corresponding to one signature is altered, another signature triggers an alarm. If they do overlap, we are able to at least derive the intersection of all occurring signatures.

*Repetitive byte-sequences.* We run into a similar problem whenever one signature matches a sequence of bytes multiple times within a file. For example, if we consider a file that contains the same signature twice, one signature always matches if we apply our derivation algorithm, as we only flip single bytes during the derivation. Consequently, it is not possible for our approach to reveal the used signature in this case. Our empirical study, however, shows that such cases are rare and we are able to retrieve byte patterns for a third of the considered signatures.

## 7  Related Work

The analysis and detection of malware is a very vivid area of research both in academia and industry. Consequently implementations of such systems are under high scrutiny, subject to security audits and target of adversarial attacks. Also, due to the widespread use of signature-based detection in commercial products, a significant body of research has particularly studied the merits and deficiencies of these systems. In the following, we discuss each of these related strains of research in detail.

*Attacking anti-virus products.* Many researchers have dealt with vulnerabilities in anti-virus products and point out implementation flaws that allow to bypass defensive mechanisms or hijack execution [2, 20, 30, 45]. Over the past few months Ormandy, for instance, called attention to several flaws in wide-spread commercial security products by ESET [31], Kaspersky [32], AVG [34] and FireEye [33] demonstrating the large attack surface such systems expose. Our method in contrast does not rely on implementation flaws, but addresses a conceptual issue in the use of poorly designed signatures.

Similar in spirit, Min and Varadharajan [25] make use of anti-virus products as an ally for an attack and introduce *"AV-Parmware"*, an advanced malware piggybacking on virus scanners. In particular, the device driver of the virus scanner is tricked into starting the malicious code on every boot and ensures a stealthy operation by providing the malware with the same high privileges as itself.

*Evasion of detection.* Of course, exploits in anti-virus products can also be used to bypass detection. However, also the underlying programming logic of file format parsers, unpackers or search algorithms might be used to evade detection. Jana and Shmatikov [17], for instance, describe two classes of bypasses: *Chameleon attacks* where a malware sample appears as a different type than it actually is and *werewolf attacks* that exploit differences in parsing logic of specific file types. Both, are based on a difference in "perception" between the product at the end-host and the virus scanner. A popular example in this context is the Portable Document Format (PDF): The Adobe Reader is to such an extent liberal in processing malformed documents that do not comply the specifications, that it is difficult to exactly reproduce its behavior in a security product [35, 44]. If a virus scanner's type-inference fails, there often is no alternative but reverting to static pattern-based signatures.

Moreover, methods for evading signature-based detection have been proposed and studied in various contexts [e.g., 12, 26, 36, 37, 46]. In mobile security research for instance, Zheng et al. [46] assess the robustness of virus scanners for mobile malware against simple transformations. Continuing this line of research, DroidChameleon [36] implements more advanced transformations, showing that in 2013, ten commercial Android anti-malware products can be bypassed using common obfuscation techniques. In a follow-up work, the authors additionally show that even one year later, these systems can still be bypassed with the same techniques [37].

*Deriving signatures.* In intrusion detection, testing signatures has received substantial attention [21, 28, 39, 40, 43]. To this end some authors attempt to manually [28] and automatically [21] derive signatures from such systems. In contrast to our method the latter however inspects the execution of the matching process on a binary level, while we observe the outcome of the matching completely passively but strategically modify the input.

Deriving signatures from anti-virus software on the contrary has received very little attention so far. To the best of our knowledge, the only exception is the work by Christodorescu and Jha [6] who explore possibilities of evaluating virus scanners, and in particular, provide first insights into the feasibility of signature extraction. Our approach notably extends this work by inferring more precise signatures, which not only represent simple byte sequences, but also express the relation to each other.

# 8 Conclusions

Despite several advanced detection techniques, such as behavioral blocking and packer-agnostic unpacking, most anti-virus products still rely on static signature matching as a fall-back mechanism. This signature-based detection is an effective and efficient tool, if appropriate and up-to-date signatures are available. However, the widespread use of this detection approach can also play into the hands of an attacker, if she gains access to the signatures and exploits them as an instrument for attack.

We inspect this threat and analyze the feasibility of anti-virus assisted attacks, where signatures derived from malware samples are implanted into benign network communication. Based on a novel method for deriving signatures from virus scanners, we show that a considerable amount of the deployed signatures corresponds to simple byte patterns that are not bound to particular file types or contexts. We then demonstrate how such rather carelessly designed signatures can be used by an attacker. To this end, we introduce the concept of *malicious markers* and present three scenarios in which an adversary is able to use such markers to remotely instruct a virus scanner to block or delete content on her behalf, turning the anti-virus product into an ally behind defense lines. While our study shows that simple signature matching can be a vulnerability rather than a protection, the concept of fending off attacks directly at the end host still is promising. To be effective in this task, the employed signatures however need to incorporate context and semantics of matched code to be used without side-effects. Dynamic approaches and heuristics, for instance, implicitly include these relations, such that we plead in favor of further advancing the use of more sophisticated and precise detection approaches in anti-virus products.

## Acknowledgments

The authors gratefully acknowledge funding from the German Federal Ministry of Education and Research (BMBF) under the projects APT-Sweeper (FKZ 16KIS0307) and VAMOS (FKZ 16KIS0534).

# Technische Universität Braunschweig

## Computer Science Reports since No. 2012-06

| | | |
|---|---|---|
| 2012-06 | S. Mennike | A Petri Net Semantics for the Join-Calculus |
| 2012-07 | S. Lity, R. Lachmann, M. Lochau, I. Schaefer | Delta-oriented Software Product Line Test Models - The Body Comfort System Case Study |
| 2013-01 | M. Lochau, S. Mennicke, J. Schroeter und T. Winkelmann | Extended Version of 'Automated Verification of Feature Model Configuration Processes based on Workflow Petri Nets' |
| 2013-02 | S. Lity, M. Lochau, U. Goltz | A Formal Operational Semantics of Sequential Function Tables for Model-based SPL Conformance Testing |
| 2013-03 | L. Giraldi, A. Litvinenko, D. Liu, H. G. Matthies, A. Nouy | To be or not to be intrusive? The solution of parametric and stochastic equations – the "plain vanilla" Galerkin case |
| 2013-04 | A. Litvinenko, H. G. Matthies | Inverse problems and uncertainty quantification |
| 2013-05 | J. Rang | Improved traditional Rosenbrock–Wanner methods for stiff ODEs and DAEs |
| 2013-06 | J. Koslowski | Deterministic single-state 2PDAs are Turing-complete |
| 2014-01 | B. Rosić, J. Diekmann | Stochastic Description of Aircraft Simulation Models and Numerical Approaches |
| 2014-02 | M. Krosche, W. Heinze | A Robustness Analysis of a Preliminary Design of a CESTOL Aircraft |
| 2014-03 | J. Rang | Apdative timestep control for fully implicit Runge–Kutta methods of higher order |
| 2014-04 | S. Mennicke, J.-W. Schicke-Uffmann, U. Goltz | Free-Choice Petri Nets and Step Branching Time |
| 2014-05 | A. Martens, C. Borchert, T. O. Geissler, O. Spinzyck, D. Lohmann, R. Kapitza | Exploiting determinism for efficient protection against arbitrary state corruptions |
| 2014-06 | J. Rang | An analysis of the Prothero–Robinson example for constructing new adaptive ESDIRK methods of order 3 and 4 |
| 2014-07 | J. Rang, R. Niekamp | A component framework for the parallel solution of the incompressible Navier-Stokes equations with Radau-IIA methods |
| 2014-08 | J. Rang | The Prothero and Robinson example: Convergence studies for Runge–Kutta and Rosenbrock–Wanner methods |
| 2014-09 | J. Wayetens, B. V. Rosic | Comparison of deterministic and probabilistic approaches to identify the dynamic moving load and damages of a reinforced concrete beam |
| 2014-10 | B. Rosic, J. Sykora, O. Pajonk, A. Kucerova, H. G. Matthies | Comparison of Numerical Approaches to Bayesian Updating |
| 2016-01 | M. Kowal, S. Ananieva, T. Thüm | Explaining Anomalies in Feature Models |
| 2016-02 | D. Arp, E. Quiring, C. Wressnegger und K. Rieck | Bat in the Mobile: A Study on Ultrasonic Device Tracking |
| 2016-03 | C. Wressnegger, K. Freeman, F. Yamaguchi und K. Rieck | From Malware Signatures to Anti-Virus Assisted Attacks |

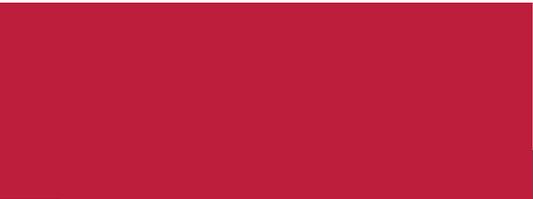

Technische Universität Braunschweig
Institute of System Security
Rebenring 56
38106 Braunschweig
Germany